\documentclass[11pt,citesort]{article}

\usepackage{geometry}
\usepackage{axodraw}
\newcommand{\ice}[1]{\relax}
\usepackage{graphicx}
\usepackage{epsfig}
\usepackage{latexsym}
\usepackage{amsmath}
\usepackage{amsfonts}
\usepackage{amsxtra}
\usepackage{cite}

 \newcommand{\alinea}{\hspace*{\parindent}}

\def\krto{ {\,\,\lower .8ex\hbox {$\longrightarrow \atop k \rightarrow 0$}\,\,}}
\def\Section#1{\section{#1}\hspace{\parindent}}

\def\alinea{\hspace{\parindent}}
\def\bea{\begin{eqnarray} }
\def\beq{\begin{eqnarray} }

\def\eea{\end{eqnarray}}
\def\eeq{\end{eqnarray}}
\def\eq#1{eq.~(\ref{#1})}

\def\assymto#1{\mbox{\raisebox{-1.2ex}[0.ex][1.6ex]{$\stackrel{\simeq}{\scriptscriptstyle #1}$}}}

\newcommand{\ghostSD}{\begin{picture}(150,25)(0,0)
\SetWidth{1.2}
\DashArrowLine(12.5,0)(37.5,0){5}
\DashArrowLine(37.5,0)(75,0){5}
\DashLine(75,0)(112.5,0){5}
\DashArrowLine(112.5,0)(137.5,0){5}
\SetWidth{1}
\Vertex(112.5,0){2}
\GlueArc(75,0)(37.5,0,90){-4}{6}
\GlueArc(75,0)(37.5,90,180){-4}{6}
\CCirc(75,0){5}{Black}{Yellow}
\CCirc(75,37.5){5}{Black}{Yellow}
\CCirc(37.5,0){5}{Black}{Yellow}
\Text(20,-10)[l]{a,k}
\Text(50,15)[l]{d,$\nu$}
\Text(100,-10)[l]{e}
\Text(100,15)[r]{f,$\mu$}
\Text(50,-10)[l]{c,q}
\Text(120,-10)[l]{b,k}
\Text(75,48)[c]{q-k}
\end{picture}}
\newcommand{\ghostDr}{\begin{picture}(100,25)(0,0)
\SetWidth{1.2}
\DashArrowLine(12.5,0)(50,0){5}
\DashArrowLine(50,0)(87.5,0){5}
\CCirc(50,0){5}{Black}{Yellow}
\Text(12.5,-10)[l]{a}
\Text(87.5,-10)[r]{b}
\Text(50,-10)[c]{k}
\end{picture}}
\newcommand{\ghostBr}{\begin{picture}(100,25)(0,0)
\SetWidth{1.2}
\DashArrowLine(12.5,0)(87.5,0){5}
\Text(12.5,-10)[l]{a}
\Text(87.5,-10)[r]{b}
\Text(50,-10)[c]{k}
\end{picture}}

\begin{document} 
\date{}

\title{ On the IR behaviour of the Landau-gauge ghost propagator }
\author{ 
Ph.~Boucaud$^a$, 
J.P.~Leroy$^a$, 
A.~Le~Yaouanc$^a$,
J. Micheli$^a$, \\
O. P\`ene$^a$, 
J.~Rodr\'iguez-Quintero$^b$
}

\maketitle

\begin{center}
$^a$Laboratoire de Physique Th\'eorique et Hautes 
Energies\footnote{Unit\'e Mixte de Recherche 8627 du Centre National de 
la Recherche Scientifique} \\
Universit\'e de Paris XI, B\^atiment 211, 91405 Orsay Cedex,
France \\
$^b$ Dpto. F\'isica Aplicada, Fac. Ciencias Experimentales,\\
Universidad de Huelva, 21071 Huelva, Spain.
\end{center}

\begin{abstract}
We examine analytically the ghost propagator Dyson-Schwinger Equation (DSE)
in the deep IR regime and prove that a finite ghost dressing function at
vanishing momentum  is an alternative solution (solution II) to the
usually assumed divergent one (solution I).
We furthermore find that the Slavnov-Taylor identities
discriminate between these two classes of solutions and strongly support the
solution II. The latter turns out to be also preferred by lattice
simulations within numerical uncertainties.

\end{abstract}

\begin{flushright}
{\small UHU-FP/08-010}\\
{\small LPT-Orsay/08-28}\\
\end{flushright}




\Section{Introduction}
How to deal with the IR behaviour of QCD ? There are three main types of approach:
\begin{itemize}
\item  Dyson Schwinger equations (DSE) and especially the untruncated one
concerning the ghost propagator.
\item	Ward-Slavnov-Taylor identities (WSTI)
\item  Lattice QCD simulations (LQCD).
\end{itemize}

Until a few years ago, there was a clear contradiction between the
standard DSE solution and LQCD results. If we call $F(q^2)$ ($G(q^2)$)
the ghost (gluon) dressing function, the standard DSE solution (later 
labelled as solution I)  predicts that $F^2(q^2)G(q^2)$ goes to a
non-vanishing constant when $q^2 \to 0$ (see for
instance~\cite{Alkofer:2000wg}  and references therein).  LQCD indicates
on the contrary in an unambiguous  way that $F^2(q^2)G(q^2) \to 0$  when
$q^2 \to0$~\cite{IlgenGrib,Boucaud:2005ce}.
The standard solution implies~\cite{Alkofer:2008jy} also that 
$G(q^2)/q^2 $ does not diverge when $q^2\to 0$ while $F(q^2)$ diverges at least as fast as $(q^2)^{-\,1/2}$. 
Regarding  lattice QCD results, they  have long been compatible with an IR-diverging $F(q^2)$, although 
definitely at  
a much slower pace.
  This discrepancy has been
tentatively charged to different types of lattice artifacts. However
more recent LQCD data obtained in large volume simulations 
\cite{Cucchieri:2007md, Bogolubsky:2007ud} show that under those conditions
the ghost dressing function IR exponent $\alpha_F$  (assuming $F(q^2) \assymto{q^2 \to 0} (q^2)^{\alpha_F}$) 
lies in the vicinity of 0.

Now, it was proven in 
\cite{Boucaud:2005ce,Finite:2006,Boucaud:2007va} that:

\begin{itemize}
\item  there exists a second class of solutions to the DSE  (later
labelled as solution II) which implies that  $F(q^2)$ goes to a
non-vanishing constant when $q^2 \to 0$ and does not constrain
$F^2(q^2)G(q^2)$. 
\item the WSTI implies under very plausible assumptions that $F(q^2)$
goes to a non-vanishing constant when $q^2 \to 0$, which imposes the
solution II of DSE.
\end{itemize}
Thus, 
{\bf the convergence of the three methods towards a finite non-vanishing ghost dressing function is very impressive.} 
\vskip 0.3cm

Furthermore, a recent numerical study of the DSE using the LQCD gluon input
finds that both cases of solutions (I and II) are found depending on the
strong coupling constant which is a free parameter in this exercise~\cite{Boucaud:2008ji}. 
Solutions exist when the coupling constant is smaller than (or equal to) a critical value. In the general case the
solutions which  come out belong to type II, but for the critical coupling constant one finds the
solution I. It was also proved that for an appropriate coupling constant the
resulting ghost dressing function (belonging to class II) fits very well with
lattice results.  

Concerning the gluon propagator the analytic methods are not so constraining.
WSTI, under a regularity hypothesis for the longitudinal-longitudinal-transverse
gluon vertex function, predicts a divergent gluon propagator 
when $q^2 \to 0$~\cite{Boucaud:2005ce,ours-ST} while
LQCD seems to point towards a finite non-vanishing gluon propagator at 
$q^2=0$ (see, for instance, \cite{Adelaide}).
A very slow divergence of the gluon propagator, not easy to see in LQCD, might 
solve this discrepancy. 

In this paper we wish to present an analytic study of the ghost propagators of both
solutions I and  II in the deep infrared in the context of the DSE. We also  
will carefully scrutinize the relationship between DSE, WSTI and LQCD 
solutions. In section 2, the ghost propagator DSE is properly renormalised and analysed in the 
deep IR regime. The two types of solutions are obtained in section 3 and their implications 
put clearly on the table. In section 4, we discuss what WSTI tells us and section 5 is 
devoted to briefly review the LQCD results for the ghost propagator. 
We conclude in section 6. In appendix \ref{App} we show how the ghost propagator DSE that 
we exploit in the next section can be generally inferred from WSTI.


\Section{The ghost propagator Dyson-Schwinger equation}\label{revisiting}
We will examine the Dyson-Schwinger equation for the ghost
propagator (GPDSE) which can be written diagrammatically as

\vspace{\baselineskip}
\begin{small}
\bea
\left(\ghostDr\right)^{-1}%
=
\left(\ghostBr\right)^{-1}%
- 
\ghostSD %
\nonumber
\eea\end{small}%
\noindent i.e., denoting by $F^{(2)}$ (resp. $G^{(2)}$) the full ghost (resp. gluon) propagator, 

\bea\label{SD}
(F^{(2)})^{-1}_{ab}(k) &=&-\delta_{ab} k^2  \\ 
&-& g_0^2 f_{acd} f_{ebf} 
\int \frac{d^4q}{(2\pi^4) }  F^{(2)}_{ce}(q)
(i q_{\nu'}) \widetilde{\Gamma}_{\nu'\nu}(-q,k;q-k) (i k_\mu) (G^{(2)})_{\mu\nu}^{fd}(q-k), \nonumber
\eea
where $\widetilde{\Gamma}$ stands for the bare ghost-gluon vertex,

\beq
\widetilde{\Gamma}_\nu^{abc}(-q,k;q-k) \ &=& \ i g_0 f^{abc} q_\nu'  
\widetilde{\Gamma}_{\nu'\nu}(-q,k;q-k) \nonumber \\
&=&
i g_0 f^{abc} \left( \ q_\nu H_1(q,k) + (q-k)_\nu H_2(q,k) \ \right) \ ,
\label{DefH12}
\eeq
where $q$ and $k$ are respectively the outgoing and incoming ghost momenta and $g_0$ is the bare coupling constant. 
Let us now consider eq.~(\ref{SD}) at small momenta $k$. After applying the decomposition for 
the ghost-gluon vertex in eq.~(\ref{DefH12}), omitting colour indices and dividing 
 both sides by $k^2$, it reads 
\begin{equation}
\label{SD1}
\begin{split}
\frac{1}{F(k^2)} & = 1 + g_0^2 N_c \int \frac{d^4 q}{(2\pi)^4} 
\left( \rule[0cm]{0cm}{0.8cm}
\frac{F(q^2)G((q-k)^2)}{q^2 (q-k)^4} 
\left[ \rule[0cm]{0cm}{0.6cm}
\frac{(k\cdot q)^2}{k^2} - q^2  
        \right]
\ H_1(q,k)
           \right) \ .
\end{split}
\end{equation} 
It should be noticed that, because of the transversality condition, $H_2$ defined in 
eq.~(\ref{DefH12}) does not contribute for the GPDSE in the Landau gauge. 

\subsection{Renormalization of the Dyson-Schwinger equation}
\alinea The integral equation eq.~(\ref{SD1}) is written in terms of bare Green functions. It is actually meaningless unless one specifies some appropriate UV-cutoff $\Lambda$ and performs the replacements $F(k^2) \rightarrow F(k^2,\Lambda)$ \dots 
 It can be cast into a renormalized form by dealing properly with UV divergencies, {\it i.e.}
\beq
g_R^2(\mu^2) &=& Z_g^{-2}(\mu^2,\Lambda) g_0^2(\Lambda) \nonumber \\
G_R(k^2,\mu^2) &=& Z_3^{-1}(\mu^2,\Lambda) G(k^2,\Lambda) \nonumber \\
F_R(k^2,\mu^2) &=& \widetilde Z_3^{-1}(\mu^2,\Lambda) F(k^2,\Lambda) \ ,
\label{Ren}
\eeq
where $\mu^2$ is the renormalization momentum and $ Z_g, Z_3$ and $\widetilde Z_3$ the renormalization constants for the coupling constant, the gluon and the ghost respectively. $ Z_g$ is related to the ghost-gluon vertex renormalization constant (defined by
 $\widetilde{\Gamma}_R=\widetilde Z_1 \Gamma_B$) through $ Z_g= \widetilde{Z_1} (Z_3^{1/2}\,\widetilde Z_3)^{-1}$. Then  
Taylor's well-known non-renormalization theorem, which states that $H_1(q,0)+H_2(q,0)=1$ in Landau gauge and 
to any perturbative order, can be invoked to conclude that 
$\widetilde Z_1$ is finite. We recall that the renormalization point is arbitrary, except for the special value $\mu = 0$ which cannot be chosen without  a loss of generality (see, in this respect, the discussion in ref~\cite{ LvS}). Thus, 
\beq\label{SDRnS}
\frac 1 {F_R(k^ 2,\mu^2)} \ = \ \widetilde Z_3(\mu^2,\Lambda) 
+ N_C \widetilde Z_1 \ g_R^2(\mu^2) \ \Sigma_R(k^2,\mu^2;\Lambda)  
\eeq
 where
\beq
\Sigma_R(k^2,\mu^2;\Lambda) &=& \int^{q^2 < \Lambda^2} \frac{d^4 q}{(2\pi)^4} 
\nonumber \\ 
&\times&
\left( \rule[0cm]{0cm}{0.8cm}
\frac{F_R(q^2,\mu^2)G_R((q-k)^2,\mu^2)}{q^2 (q-k)^4} 
\left[ \rule[0cm]{0cm}{0.6cm}
\frac{(k\cdot q)^2}{k^2} - q^2  
        \right]
\ H_{1,R}(q,k;\mu^2) \right) \ . \nonumber \\
\label{sigma}
\eeq 
One should notice that the UV cut-off, $\Lambda$, is still required as an upper integration bound 
in eq.~(\ref{sigma}) since the integral is UV-divergent, behaving as 
$\int dq^2/q^2 (1+11 \alpha_S/(2\pi) \log{(q/\mu)}))^{-35/44}$. In fact, the 
cut-off dependence this induces in $\Sigma_R$ cancels \footnote{One can easily check that 
$\widetilde Z_3^{-1}(\mu^2,\Lambda) \Sigma_R(k^2,\mu^2;\Lambda)$ approaches  some finite limit 
as $\Lambda \to \infty$ since the ghost and gluon propagator anomalous dimensions and the
beta function verify the relation $2 \widetilde \gamma + \gamma + \beta = 0$.}  against the one 
of $\widetilde Z_3$ in the r.h.s. of eq.~(\ref{SDRnS}), in accordance with the fact that the l.h.s. does not 
depend on $\Lambda$.

Now, we will apply a MOM renormalization prescription. This means that all the Green functions 
take their tree-level value at the renormalization point and thus:
\beq
F_R(\mu^2,\mu^2) \ = \ G_R(\mu^2,\mu^2) \ = \ 1 \ .
\eeq
In the following, $H_1(q,k)$ will be approximated by a constant\footnote{This  approximation is very usually used to solve GPDSE. Notice that few lattice data are available for the 
ghost-gluon vertex. However, in a recent computation~\cite{IlgenGrib} 
of that vertex for a zero gluon momentum, $H_1(q,q)$ appears to be approximatively 
constant with respect to $q$. Of course, more data for different kinematical configurations should 
be welcome to check that approximation.} with respect to both 
momenta and, provided that $H_1(q,0)=1$ at tree-level, our MOM prescription implies 
that $H_{1,R}(k,q;\mu^2)=1$ and $\widetilde Z_1$ is a constant in terms of $\mu$.

\subsection{A subtracted Dyson-Schwinger equation} \label{subtracted}
\alinea The renormalized GPDSE, eq.~(\ref{SDRnS}), should be carefully analysed. We aim to study the 
infrared behaviour of its solutions and therefore focus our analysis on the momentum region, 
$k \ll \Lambda_{\rm QCD}$, where the IR behaviour of the dressing functions (presumably in powers of 
the momentum) is supposed to hold. One cannot forget, though, that the UV cut-off dependences in both sides 
of eq.~(\ref{SDRnS})  match only in virtue of the previously mentionned relation between the ghost and gluon propagator anomalous 
dimension and the beta function. 

However, in order  not to have to deal with the UV cut-off, we prefer  to approach the study of the GPDSE
in the following manner: 
we consider eq.~(\ref{SDRnS}) for two different scales, $\lambda k$ and $\lambda \kappa k$ 
(with $\kappa<1$ some fixed number and $\lambda$ an extra parameter that we 
shall ultimately let go to 0) and subtract them 
\beq
\frac{1}{F_R(\lambda^2 k^2,\mu^2)} - \frac{1}{F_R(\lambda^2 \kappa^2 k^2,\mu^2)}  
\ = \  
N_C \ g_R^2(\mu^2) \ \widetilde Z_1 \ 
\left( \rule[0cm]{0cm}{0.5cm} \Sigma_R(\lambda^2 k^2,\mu^2;\infty) - 
\Sigma_R(\lambda^2 \kappa^2 k^2,\mu^2;\infty) \right) \ . \nonumber \\
\label{SDRS}
\eeq
Then the integral in the r.h.s. is UV-safe, thanks to the subtraction, and the limit 
$\Lambda \to \infty$ can be explicitely taken,
\beq
\label{LamInf}
 \Sigma_R(\lambda^2 k^2,\mu^2;\infty) - 
\Sigma_R(\lambda^2 \kappa^2 k^2,\mu^2;\infty) 
&=&  \int \frac{d^4 q}{(2\pi)^4} 
\left( \rule[0cm]{0cm}{0.8cm}
\frac{F(q^2,\mu^2)}{q^2} \left(\frac{(k\cdot q)^2}{k^2}-q^2\right) \right. 
\nonumber \\ 
 &\times& \left. 
\left[ \rule[0cm]{0cm}{0.6cm}
\frac{G((q-\lambda k)^2,\mu^2)}{\left(q-\lambda k\right)^4} -  
(\lambda \to \lambda \kappa)
\rule[0cm]{0cm}{0.6cm} \right]
\rule[0cm]{0cm}{0.8cm} \right) \ .
\eeq
An accurate analysis of eq.~(\ref{SDRS}) requires, in addition,  to cut the 
integration domain of eq.~(\ref{LamInf}) into two pieces by introducing 
some new scale $q_0^2$ ($q_0$, typically of the order of $\Lambda_{\it QCD}$, 
is a momentum scale below which the deep IR power behaviour is a 
good approximation),
\beq\label{q0}
\Sigma_R(\lambda^2 k^2,\mu^2;\infty) - \Sigma_R(\lambda^2 \kappa^2 k^2,\mu^2;\infty) 
\ = \ 
I_{\rm IR}(\lambda) \ + \ I_{\rm UV}(\lambda)
\eeq
where $I_{\rm IR}$ represents the integral in eq.~(\ref{LamInf}) over $q^2 < q_{0}^2$ and 
$I_{\rm UV}$ over $q^2 > q_{0}^2$.
%
%
Only the dependence on $\lambda$ is written explicitly because we  shall let it go to zero 
with $k$, $\kappa$ and $\mu^2$ kept fixed. 
The relevance of the $q_0^2$  scale stems from the drastic difference  between the IR and UV behaviours of the integrand.  In particular, for $(\lambda k)^2 \ll q_{0}^2$, the 
following infrared power laws, 
\beq\label{dress}
F_{\rm IR}(q^2,\mu^2) &=& A(\mu^2) \left( q^2 \right)^{\alpha_F}
\nonumber \\
G_{\rm IR}((q-\lambda k)^2,\mu^2) &=& B(\mu^2) \left( (q-\lambda k)^2 \right)^{\alpha_G} \ ,
\eeq
will be applied for both dressing functions in $I_{\rm IR}$.

Now, $I_{\rm IR}$ is infrared convergent if :
\begin{eqnarray}
\label{cond_ward}
\alpha_F &>& -2 \qquad{\rm IR\; convergence \; at}\; q^2 = 0 \nonumber \\
\alpha_G &>& 0 \qquad{\rm IR\; convergence\; at}\; (q-k)^2 = 0 \; 
{\rm and} \;(q-\kappa k)^2 = 0
\end{eqnarray}
We shall suppose in the following that these conditions are verified. We then obtain, 
performing the change of variable $q\to \lambda q$ :
\bea
I_{\rm IR}(\lambda) 
&\simeq& 
\left(\lambda^2 \right)^{(\alpha_F + \alpha_G )} A(\mu^2) B(\mu^2) 
\displaystyle \int^{q^2 < 
\frac{q_{0}^2}{\lambda^2}} \frac{d^4 q}{(2\pi)^4} \ \
(q^2)^{\alpha_F-1} \          
\left( 
\frac{(k\cdot q)^2}{k^2}-q^2
\right)  
\nonumber \\ 
& & \times 
\left[ 
  \left(  (q-k)^2 \right)^{\alpha_G  -2} -
  \left((q-\kappa k)^2\right)^{\alpha_G -2} 
\right]
\label{I1}
\eea
The point we have to keep in  mind is the fact that
the upper bound of the integral goes to infinity when $\lambda \to 0$. 
This potentially induces a dependence on $\lambda$ whose interplay with 
the behaviour explicitly shown in (\ref{I1}) we must check.
In this limit, the convergence of the integral depends on the asymptotic 
behaviour of the whole integrand for large $q$. In particular, the 
leading contribution of the square bracket in  eq.~(\ref{I1}) behaves 
as 
\bea\label{expa}
\left[ (k - q)^2 \right]^{\alpha_G -2} - 
\left[ (\kappa k - q)^2 \right]^{\alpha_G -2} &\simeq& (q^2)^{\alpha_G-2} \ (\alpha_G-2) (1-\kappa)
\\  
& \times & \left[ - 2 \ \frac{q \cdot k}{q^2} \ + \ (1+\kappa) \left( \frac {k^2}{q^2} 
+ 2 (\alpha_G-3) \frac{(q \cdot k)^2}{q^4} \right) \right] \ ;
\nonumber 
\eeq
where we expand up to $(k^2/q^2)$-terms because those in $q \cdot k$,  being odd under
 $q_\mu \to -q_\mu$, give a null contribution under the angular integration in eq.~(\ref{I1}). 
Thus, provided that the conditions~(\ref{cond_ward}) are satisfied so that  $I_{\rm IR}$ 
is convergent when $q \to 0$ (or $q \to k$), its asymptotics for small $\lambda$ is 
\bea\label{I1a}
I_{\rm IR}(\lambda) \sim \lambda^{2(\alpha_G+\alpha_F)} \int_\epsilon^{q_0/\lambda} dq \ 
q^{2(\alpha_F+\alpha_G)-3} \ ;
\eea
where $\epsilon$, the lower limit of the integral, is a small cut-off that  
avoids any possible spureous singularity appearing after expanding 
in eq.~(\ref{I1}) for large $q$.   
Thus, if $\alpha_F+\alpha_G  < 1$, the asymptotic behaviour of $I_{\rm IR}$ is given by 
the power on $\lambda$ in front of the integral in eq.~(\ref{I1a}), since the 
integral itself will remain finite when $\lambda \to 0$. If $\alpha_F+\alpha_G  = 1$, the integral 
diverges logarithmically as $\lambda$ vanishes. Otherwise, one can change 
the integration variable back, $q \to q/\lambda$, to get the leading power on $\lambda$ 
multiplying again a finite integral on the momentum $q$, 
\bea
I_{\rm IR}(\lambda) \sim \left\{ 
\begin{array}{ll} 
\displaystyle 
\lambda^{2(\alpha_G+\alpha_F)}
 & \mbox{\rm if} \ \ \alpha_G+\alpha_F  < 1\\
\displaystyle 
\lambda^2 \ \ln{\lambda} &  \mbox{\rm if} \ \ \alpha_G+\alpha_F  = 1 \\
\displaystyle 
\lambda^2 & 
\mbox{\rm if}   \ \ \alpha_G+\alpha_F  > 1
\end{array} 
\right.
\label{cond_alpha_F_G}
\eea
We have assumed that $H_1$ is constant when varying all the momenta but (\ref{cond_alpha_F_G}) 
remains true if one only assumes that $H_1$  behaves ``regularly'' for $q^2, k^2 \le q_{0}^2$
(i.e. is free of singularities or, at least, of any singularity worse than logarithmic).

Let us now consider  $I_{\rm UV}$. Its  dependence on $\lambda$, which is explicit in the 
factor inside the square bracket of eq.~(\ref{LamInf}),  
should clearly be even in $\lambda$ : any odd power of  $\lambda$  would imply an odd 
power of $q \cdot k$ whose angular integral is zero.  
Since the integrand is identically zero at $\lambda = 0$ and the integral is ultraviolet 
convergent, it is  proportional to $\lambda^2$ (unless some accidental cancellation 
forces it to behave as an even higher  power of  $\lambda$). Thus, in all the cases, 
the leading behaviour of $I_{\rm IR}+I_{\rm UV}$, as $\lambda$ vanishes, is given by $I_{\rm IR}$ 
in eq.~(\ref{cond_alpha_F_G}). The subtracted renormalised GPDSE reads 
for $\alpha_G+\alpha_F \leq 1$ as:
\beq
\frac{1}{F_R(\lambda^2 k^2,\mu^2)} - \frac{1}{F_R(\lambda^2 \kappa^2 k^2,\mu^2)}  
\ \simeq \  
N_C \ g_R^2(\mu^2) \ \widetilde Z_1 \ 
I_{\rm IR}(\lambda) \ ,
\label{SDRSF}
\eeq
for small $\lambda$.

\subsection{The ghost-loop integral}
\alinea A more quantitative analysis than the one presented in the preceding section can be done 
if we compute exactly the integral $I_{\rm IR}(\lambda)$, defined 
in \eq{I1}, which gives the contribution of the
 ghost loop to the renormalised GPDSE \eq{SDRSF}. If $\alpha_F+\alpha_G < 1$ it is possible to  
perform analytically the integral and to find a compact expression for it. In this case, one can write
\beq
I_{\rm IR}(\lambda) 
&\simeq& 
A(\mu^2) B(\mu^2) \ \left(\lambda^2 \right)^{(\alpha_F + \alpha_G )}
\left( \Phi(k;\alpha_F,\alpha_G) - \Phi(\kappa k;\alpha_F,\alpha_G) 
\rule[0cm]{0cm}{0.5cm} 
\right) 
\label{I2}
\eeq
where $A(\mu^2)$ and $B(\mu^2)$ were defined in \eq{dress} and
\beq
\Phi(k;\alpha_F,\alpha_G) = 
\int \frac{d^4 q}{(2\pi)^4} \ \
(q^2)^{\alpha_F-1} \ \left(  (q-k)^2 \right)^{\alpha_G  -2}         
\left( 
\frac{(k\cdot q)^2}{k^2}-q^2
\right)  \ ,
\label{BigPhi}
\eeq
provided that $\Phi(k;\alpha_F,\alpha_G)$ is not singular to let the subtraction inside the bracket and 
the integral operator in \eq{I2} commute with each other.
Then, following~\cite{Bloch:2003yu}, we define
\beq
f(a,b) \ &=& \ \frac {16 \pi^2} {(k^2)^{2+a+b}} \ \int \frac{d^4 q}{(2\pi)^4} 
(q^2)^a
\left(  (q-k)^2 \right)^b \nonumber \\
&=&  
\frac{\Gamma(2+a) \Gamma(2+b) \Gamma(-a-b-2)}
{\Gamma(-a) \Gamma(-b) \Gamma(4+a+b)} \ ,\label{fgammas}
\eeq
and obtain
\beq\label{smallphi}
\Phi(k;\alpha_F,\alpha_G) \ = \ 
\frac{(k^2)^{\alpha_F+\alpha_G}}{16 \pi^2} \ \phi(\alpha_F,\alpha_G)
\eeq
where
\beq
\phi(\alpha_F,\alpha_G)  &=& 
-\frac 1 2 \left( 
  f(\alpha_F,\alpha_G-2) + f(\alpha_F,\alpha_G-1) 
  + f(\alpha_F -1,\alpha_G-1)
\right) \nonumber \\
&+& 
 \frac 1 4 \left(
  f(\alpha_F-1,\alpha_G-2) + f(\alpha_F-1,\alpha_G) 
  + f(\alpha_F+1,\alpha_G-2)
\right)  \ .\label{phifs}
\eeq
Thus, if $\alpha_F+\alpha_G < 1$,
\beq\label{IR<1}
I_{\rm IR}(\lambda) \ \simeq \ \frac {A(\mu^2) B(\mu^2)}{16 \pi^2} (\lambda^2 k^2)^{\alpha_F+\alpha_G}
(1-\kappa^{2(\alpha_F+\alpha_G)}) \ \phi(\alpha_F,\alpha_G) \ .
\eeq

On the other hand, we know from \eq{cond_alpha_F_G} that $I_{\rm IR}$ 
diverges logarithmically as $\lambda$ goes to zero
if $\alpha_F+\alpha_G=1$. In fact, since \eq{IR<1} is a reliable result for any $\alpha_F+\alpha_G < 1$ how close it may be 
 to 1, such a divergence appears as a pole of a Gamma function of $\phi(\alpha_F,\alpha_G)$ 
in \eq{smallphi}. We will now compute the leading asymptotic behavior of $I_{\rm IR}$ as $\lambda \to 0$ when 
$\alpha_F+\alpha_G=1$.

In that case, after performing in \eq{I2} the expansion \eq{expa} and neglecting the term odd in $q_\mu \to -q_\mu$, 
one finds for the leading contribution 
\beq\label{IR=1}
I_{\rm IR}(\lambda) & \simeq &
- k^2 (1-\kappa^2) \frac {2 A(\mu^2) B(\mu^2)}{(2 \pi)^3} \ 
\lambda^2 \int^{q_0/\lambda} dq \ q^{2(\alpha_F+\alpha_G)-3}
\nonumber \\
&\times& \int_0^\pi d\theta \ {\rm sin}^4\theta 
\left( \alpha_G-2 + 2(\alpha_G-3)(\alpha_G-2) {\cos}^2\theta 
\rule[0cm]{0cm}{0.5cm}
\right)
\nonumber \\
& \simeq &
k^2 (1-\kappa^2) \frac {A(\mu^2) B(\mu^2)}{32 \pi^2} \ \alpha_G (\alpha_G-2) \lambda^2 \ln{\lambda} \ .
\eeq
We do not specify the lower bound of the integral over $q$ in \eq{IR=1} because it necessarily 
contributes as a subleading term, once the ghost-loop integral is required to be IR safe.

Finally, if $\alpha_F+\alpha_G > 1$, the leading contribution for $I_{\rm IR}(\lambda)$ 
as $\lambda$ vanishes can be computed 
after performing back the change of integration variable, $q \to q/\lambda$, in \eq{I1}. The first 
even term in \eq{expa} dominates again the expansion after integration, but now it does 
not diverge. Then, if we procceed as we did in \eq{IR=1}, we obtain
\beq\label{IR>1}
I_{\rm IR}(\lambda) \ \simeq \ - \frac{\alpha_G (\alpha_G-2)}{\alpha_F+\alpha_G-1} \
\frac{(q_0^2)^{\alpha_F+\alpha_G-1}}{64 \pi^2} \ A(\mu^2) B(\mu^2) \ k^2 \lambda^2 (1-\kappa^2) \ ,
\eeq
for small $\lambda$ and $\alpha_G+\alpha_F > 1$. It should be noticed that $I_{\rm IR}$ in \eq{IR>1} 
depends on the additional scale $q_0$ introduced in \eq{q0} to separate IR and UV integration domains.
In fact, if one takes $q_0 \to \infty$, $I_{\rm IR}$ diverges. This means that, when 
$\alpha_F+\alpha_G > 1$, the behaviour of the IR power laws  hampers their use for all momenta in the integral. The 
finiteness of the ghost-loop integral of the subtracted GPDSE can only be recovered  after taking into account the 
UV logarithmic behaviour for large-momenta 
dressing functions~\footnote{The scale $q_0$ being of the order of $\Lambda_{\rm QCD}$, power laws 
for $\alpha_F+\alpha_G > 1$ cannot be then solutions of the GPDSE in the MR truncation scheme corresponding to 
$\Lambda_{\rm QCD} \to \infty$ (see, for instance,~\cite{Bloch:2003yu}). The same argument holds also for 
$\alpha_F+\alpha_G=1$, because the ghost-loop integral in \eq{IR=1} diverges as $\lambda \to 0$ for any $q_0$ fixed 
as well as for $q_0 \to \infty$ for any fixed $\lambda$.}. Furthermore, $I_{\rm UV}$, also behaving as $\lambda^2$, 
should be also added in r.h.s. of \eq{SDRSF} in order to write the renormalised GPDSE.
Thus, the dependence on $\lambda$ but not the factor in 
front of it can be inferred 
from the GPDSE with only the information of the assymptotics for small-momentum dressing functions.

\section{The infrared analysis of GPDSE solutions}
\alinea The starting point for the infrared analysis will be the \eq{SDRSF} for small $\lambda$, where we will try to 
make the dependences on $k,\kappa$ and $\lambda$ of the two sides match each other. 

\subsection{The case $\alpha_F \neq 0$ (solution I)}
\alinea We will first study the case $\alpha_F \neq 0$. Then, the l.h.s. of \eq{SDRSF} can be 
expanded for small $\lambda$ as
\beq
\frac{1}{F_R(\lambda^2 k^2,\mu^2)} - \frac{1}{F_R(\lambda^2 \kappa^2 k^2,\mu^2)}  
\simeq 
\left( 1-\kappa^{-2 \alpha_F} \right) \
\frac{\left( \lambda^2 k^2 \right)^{-\alpha_F}}{A(\mu^2)} 
\label{lhsSDRSF}
\eeq
and we will obtain from \eq{SDRSF}:
\beq
N_C \ g_R^2(\mu^2) \ \widetilde Z_1  A(\mu^2) \
\frac{I_{\rm IR}(\lambda)}
{\left( 1-\kappa^{-2 \alpha_F} \right) \left( \lambda^2 k^2 \right)^{-\alpha_F} }
\ \simeq \ 1 \ ,
\label{SDRS2}
\eeq
where the dependences on $k,\kappa$ and $\lambda$ of the  numerator and  the denominator
should cancel against each other. Using for  $I_{\rm IR}$ the form given in after \eq{cond_alpha_F_G}, 
we find three possible situations:

\begin{itemize}

\item If $\alpha_G+\alpha_F > 1$,  applying \eq{IR>1} in \eq{SDRS2}, we are led to the conclusion that 
only \fbox{$\alpha_F=-1$}~(and $\alpha_G  > 2$) satisfies this last 
equation and could be an IR solution for GPDSE. 
However, such  a solution appears to be in a clearcut contradiction with the current lattice simulations.

\item If $\alpha_G+\alpha_F=1$, there is no possible solution because the logarithmic 
behaviour of $I_{\rm IR}$ in \eq{IR=1} cannot be compensated by the powerlike one in the denominator 
of \eq{SDRS2}.  

\item If $\alpha_G+\alpha_F < 1$, \eq{IR<1} combined with \eq{SDRS2} implies the familiar relation   
\fbox{$2 \alpha_F+\alpha_G=0$} and we have then:
\beq
N_C \ g_R^2(\mu^2) \ \widetilde Z_1  \frac{(A(\mu^2))^2 B(\mu^2)}{16 \pi^2} \
\phi\left(-\frac{\alpha_G} 2,\alpha_G\right)
\ \simeq \ 1 \ ,
\label{SDRS2<1}
\eeq

\end{itemize}

An immediate consequence of this last condition is the freezing of the running coupling 
constant at small momentum. If the renormalization point, $\mu$, is arbitrarily chosen 
to be very small in order that the dressing functions observe the power laws 
at $k^2=\mu^2$, one obtains $A(\mu^2)=\mu^{-2 \alpha_F}$ and $B(\mu^2)=\mu^{-2 \alpha_G}$. 
Eq.~(\ref{SDRS2<1}) then reads
\beq\label{SDRS2<1deep}
N_C \ g_R^2(\mu^2) \ \widetilde Z_1 \  \phi\left(-\frac{\alpha_G} 2,\alpha_G\right) \ 
\simeq 16 \pi^2 \ ,
\eeq
and should be satisfied for any small value of 
$\mu$. Consequently, it should remain exact as $\mu \to 0$ and provides the small-momentum 
limit of the running coupling (which is independent of the infrared constants for 
ghost and gluon dressing functions). 

In particular, if $\alpha_G=1$, one has $\phi(-1/2,1)=8/5$ and thus
\beq\label{SDRS2<1aG=1}
N_C \ g_R^2(\mu^2) \ \widetilde Z_1 \  
\simeq 10 \pi^2 \ ,
\eeq

\subsection{The case $\alpha_F=0$ (solution II)}
\alinea The case $\alpha_F=0$ is particular in  that the leading contributions to the two occurrences of $F$ in the l.h.s of 
eq.~(\ref{SDRSF}) cancel against each other. 
We have then to go one step further, taking into account the subleading terms.   Defining $\widetilde F_{IR}$ by means of $ F_{IR}(q^2,\mu^2)=A(\mu^2) + \widetilde F_{IR}(q^2,\mu^2)$ we rewrite the  l.h.s of eq.~(\ref{SDRS}) as $ -(\widetilde F_{IR}(\lambda^2 k^2,\mu^2) -\widetilde  F_{IR}(\lambda^2 \kappa^2k^2,\mu^2))/A^2(\mu^2)$ and use the known IR behaviour of $I_{\rm IR}(\lambda)$ from eq.\ref{cond_alpha_F_G})) in the r.h.s. of eq.~(\ref{SDRSF}) to get 
\beq\label{FIR2}
F_{\rm IR}(q^2,\mu^2) \ = \ 
\left\{ 
\begin{array}{ll}
A(\mu^2) + A_2(\mu^2) q^2 \ln{q^2} & {\rm if~} \alpha_G=1 \\
A(\mu^2) + A_2(\mu^2) (q^2)^{\alpha_F^{(2)}} & {\rm otherwise} \ .
\end{array}
\right.
\eeq 
Furthermore, not only the subleading functional behaviour of the dressing function 
can be constrained but also the coefficient $A_2$ in \eq{FIR2}. In fact, if we
plug this equation into the l.h.s. of eq.~(\ref{SDRSF}) and expand we obtain :
\beq
- \frac{(A(\mu^2))^2}{A_2(\mu^2)} \ N_C \ g_R^2(\mu^2) \ \widetilde Z_1 \
I_{\rm IR}(\lambda) \ 
\simeq \ 
\left\{
\begin{array}{ll} 
k^2 (1-\kappa^2) \lambda^2 \ln{\lambda^2} & {\rm if~} \alpha_G=1 \\
(\lambda^2 k^2)^{\alpha_F^{(2)}} (1-\kappa^{2\alpha_F^{(2)}}) & {\rm otherwise} \ ,
\end{array}
\right.
\label{SDRS3}
\eeq
Let us consider now in more detail the three possible cases.
\begin{itemize}

\item If $\alpha_G < 1$, we obtain from eqs.~(\ref{IR<1},\ref{SDRS3}) that 
\fbox{$\alpha_F^{(2)}=\alpha_G$}~. Then, 
\beq
- \frac{(A(\mu^2))^3 B(\mu^2)}{A_2(\mu^2)} \ N_C \ g_R^2(\mu^2) \ \widetilde Z_1 \
\phi(0,\alpha_G) 
\simeq \ 
16 \pi^2 \ ,
\label{SDRS3<1}
\eeq
where, according to eqs.~(\ref{fgammas},\ref{phifs}) $\phi(0,\alpha_G)$ is given by 
\beq\label{solphi}
\phi(0,\alpha_G) \ = \ \frac 3 {2 \alpha_G (\alpha_G+1)(\alpha_G+2)(1-\alpha_G)}
\eeq

\item Similarly if $\alpha_G=1$, \eq{IR=1} applied to \eq{SDRS3} leads to
\beq
\frac{(A(\mu^2))^3 B(\mu^2)}{A_2(\mu^2)} \ N_C \ g_R^2(\mu^2) \ \widetilde Z_1 \
\simeq \ 
64 \pi^2 \ .
\label{SDRS3=1}
\eeq

\item At last, if $ \alpha_G > 1$, eqs. (\ref{IR>1}) and  (\ref{SDRS3}) imply:~\fbox{$\alpha_F^{(2)}=1$}~. 
{\it i.e.}, a ghost dressing function which behaves quadratically for small momenta,
In this case, however, as already said the ghost loop cannot be evaluated using the IR power laws  over the whole integration range and it is therefore not possible to solve the GPDSE  consistently, nor even to determine the small-momentum behaviour 
of the dressing functions, without matching appropriately those power laws to the UV perturbative 
formulas. Thus, we are not able to derive a constraint for the next-to-leading coefficient, $A_2(\mu^2)$.

\end{itemize}

In summary, the GPDSE admits IR solutions with $\alpha_F=0$ and any $\alpha_G > 0$, provided that 
\beq\label{solghost}
F_{\rm IR}(q^2,\mu^2) \ = \ 
\left\{
\begin{array}{ll}
A(\mu^2) \left( 1 -  \phi(0,\alpha_G) \displaystyle \frac{ \widetilde{g}^2(\mu^2)}{16 \pi^2} A^2(\mu^2) B(\mu^2)
(q^2)^{\alpha_G} \right) & \alpha_G < 1 \\
A(\mu^2) \left( 1 + \displaystyle \frac{\widetilde{g}^2(\mu^2)}{64 \pi^2} A^2(\mu^2) B(\mu^2) \ 
q^2 \ln{q^2} \right) & \alpha_G = 1 \\
\rule[0cm]{0cm}{0.5cm} A(\mu^2) + A_2(\mu^2) q^2 & \alpha_G > 1
\end{array}
\right.
\eeq
where $\widetilde{g}^2(\mu^2)=N_C \ g_R^2(\mu^2)  \widetilde{Z_1}$ and $\phi(0,\alpha_G)$ is given in \eq{solphi}. 
The gluon dressing function is supposed to behave as indicated in \eq{dress}. 
In particular for $\alpha_G=1$, the gluon propagator 
takes a finite (and non-zero) value at zero momentum, $B(\mu^2)$, 
after applying MOM renormalisation prescription at $q^2=\mu^2$.

\section{The ghost-gluon and three-gluon Ward-Slavnov-Taylor identity}\label{SlavTayl}
\alinea In the previous section, we have analysed the infrared behaviour of GPDSE solutions and found that the 
ghost dressing function  can either diverge at vanishing momentum 
($\alpha_F=-\alpha_G/2$ with $\alpha_G > 0$) or give a finite value 
($\alpha_F=0$ with any $\alpha_G > 0$). As appendix \ref{App} shows, the GPDSE can be 
derived from the general Ward-Slavnov-Taylor equation~\cite{Taylor}.
We will now invoke the Ward-Slavnov-Taylor identity (WSTI) for general 
covariant gauges relating the 3-gluon, $\Gamma_{\lambda\mu\nu}(p,q,r)$, 
and ghost-gluon vertices, 
\begin{equation}
\label{STid}
\begin{split}
p_\lambda\Gamma_{\lambda \mu \nu} (p, q, r) & =
\frac{F(p^2)}{G(r^2)} (\delta_{\rho\nu} r^2 - r_\rho r_\nu) \widetilde{\Gamma}_{\rho\mu}(r,p;q) 
\\ & -
\frac{F(p^2)}{G(q^2)} (\delta_{\rho\mu} q^2 - q_\rho q_\mu) \widetilde{\Gamma}_{\rho\nu}(q,p;r) \ .
\end{split}
\end{equation}
to shed some light on that matter~\cite{ours-ST}. 
Using for the ghost-gluon 
vertex the  general decomposition\footnote{We work of course on the energy-momentum shell, 
so that the relation $p + q + r \equiv 0$ holds}~\cite{Ball:1980ax}
\beq\label{Def2}
 \widetilde{\Gamma}_{\nu\mu}(p,q;r) &=& 
\delta_{\nu\mu} \ a(p,q;r) \ - \ r_\nu q_\mu \ b(p,q;r) 
\ + \ p_\nu r_\mu \ c(p,q;r)\nonumber
\\ 
&+& r_\nu p_\mu \ d(p,q;r) \ + \  p_\nu p_\mu \ e(p,q;r) \ ,
\eeq
and multiplying by $r_\nu$  both sides of \eq{STid}, one obtains:
\beq\label{STid2}
r_\nu p_\lambda \Gamma_{\lambda\mu\nu}(p,q,r) = 
\frac{F(p^2)}{G(q^2)} X(q,p;r) \ 
\left[ (q \cdot r) q_\mu - q^2 r_\mu  \right]  \ ;
\eeq
where 
\beq\label{combi}
X(q,p;r) \ = \ a(q,p;r)- (r \cdot p) \ b(q,p;r) + (r \cdot q) \ d(q,p;r) \ .
\eeq
Since the vertex function, $\Gamma$, in the l.h.s. of \eq{STid2} is antisymmetric under $p \leftrightarrow r$
and $\lambda \leftrightarrow \nu$, one can then conclude that\cite{ours-ST,Chetyrkin:2000dq}:
\begin{equation}\label{Ghexp}
F(p^2) X(q,p;r)= F(r^2) X(q,r;p) \ .
\end{equation} 
This last result is a compatibility condition required for the WSTI to be satisfied that 
does not involve the 3-gluon vertex and implies a strong correlation between the infrared 
behaviours of the ghost-gluon vertex and the ghost propagator. 
Now, under the only additional hypothesis that those scalars of the ghost-gluon vertex 
decomposition in \eq{Def2} contributing to the scalar function $X$ defined 
in \eq{combi} are regular\footnote{Note also that, for our purposes, it will actually be enough to 
restrict, and not forbid, the possible presence of singularities in the scalar coefficient 
functions provided that they could be compensated by kinematical zeroes 
stemming from the tensors.} 
when one of their arguments goes to zero while the others 
are kept non-vanishing, one can consider the small $p$ limit in \eq{Ghexp} and obtain:
\beq
F(p^2) X(q,0;-q)= F(q^2) X(q,-q;0) +{\cal O}(p^2)\ 
\eeq
This has to be true for {\bf any} value of $q$, which implies $F(p^2)$  {\bf goes to some finite and non-zero value when $p$ goes to zero}, since neither $X(q,0;-q)$ nor $X(q,-q;0)$ are presumably zero for all values of $q$. Rephrased in terms of infrared exponents, the latter argument  implies that  $\alpha_F=0$.

To reach the above conclusions we did not appeal to the properties of the  3-gluon vertex, apart from the symmetry under the exchange of gluon legs. If one assumes in addition that the longitudinal 
part of the 3-gluon vertex also behaves regularly when anyone of its arguments goes to $0$, the 
others being kept non-vanishing, a divergent gluon propagator at vanishing momentum will be 
implied~\cite{Boucaud:2005ce,Boucaud:2007va,ours-ST}. 
Of course, as far as it involves a vertex with longitudinal gluons which have not been very extensively studied, 
this last conclusion is not as clean as the previous one about the ghost dressing (according to authors of
ref.~\cite{Alkofer:2008jy} a soft kinematical singularity appears for the landau-gauge 3-gluon vertex,
however it does not concern our proof relying on the regularity of the longitudinal-longitudinal-transverse 3-gluon vertex).

In ref.~\cite{ours-ST}, we showed that only a very mild divergence, for example of logarithmic type, could be compatible 
with current LQCD results for the gluon propagator. 
The IR analysis of the previous section can be straightforwardly extended  to this case 
by generalizing
\beq
G_{\rm IR}(q^2,\mu^2) \ = \ B(\mu^2) \left(q^2\right)^{\alpha_G} \log^{\nu}\left({\frac 1 {q^2}}\right) \ ,
\eeq
the effect of which is to modify eq.~(\ref{solghost}) with
 
\beq
\label{solghostlog}
F_{\rm IR}(q^2,\mu^2) \ = \ 
\left\{
\begin{array}{ll}
A(\mu^2) \left( 1 -  \phi(0,\alpha_G) \displaystyle \frac{ \widetilde{g}^2(\mu^2)}{16 \pi^2} A^2(\mu^2) B(\mu^2)
(q^2)^{\alpha_G} \log^{\nu}{(q^{-2})} \right) & \alpha_G < 1 \\
A(\mu^2) \left( 1 - \displaystyle \frac{\widetilde{g}^2(\mu^2)}{(\nu+1) 64 \pi^2} \ A^2(\mu^2) B(\mu^2) \ 
q^2  \log^{(\nu + 1)}{(q^{-2})}\right) & \alpha_G = 1 \\
\rule[0cm]{0cm}{0.5cm} A(\mu^2) + A_2(\mu^2) q^2 \log^{\nu}{(q^{-2})} & \alpha_G > 1
\end{array}
\right.
\eeq
where only the power of the logarithm is then modified.

Sticking now to the case where $\alpha_F$ is zero (for the reasons explained above) 
and $\alpha_G$ is 1 (as suggested by the lattice results) we are left with 
%
%
\beq\label{FIR-lat}
F_{\rm IR}(q^2,\mu^2) \ = \ F_{\rm IR}(0,\mu^2) \left( 1 - \frac{\widetilde{g}^2(\mu^2)}{(\nu + 1)64 \pi^2} \ 
F_{\rm IR}(0,\mu^2)^2 B(\mu^2) \ q^2 \log^{(\nu + 1)} {\left(\frac{M^2}{q^2}\right)}  \right)
\ ,
\eeq
according to whether there are logarithmic corrections to the gluon propagator ($\nu \neq 0$) or 
not ($\nu=0$). Here, $M$ is some scale which is out of the scope of the IR analysis we performed in the 
previous section and, if $\nu=0$,  $B(\mu^2)=G_{\rm IR}^{(2)}(0,\mu^2)$ is the gluon propagator 
at zero momentum.

%
%




\Section{Ghost propagator from LQCD}
The theoretical study by Zwanziger~\cite{ZWA94} of the Faddeev-Popov 
operator on the lattice in Landau gauge triggered the first Lattice simulation of the ghost propagator~\cite{SUM96} in SU(2) and SU(3) 
gauge theories and the subsequent activity which, mainly for technical reasons, 
 was mostly dedicated to the SU(2) lattice gauge theory in the
infrared region. It was only in the last few years that several studies of the SU(3) ghost
propagator focused on its infrared region and
the Gribov copy problem~\cite{IlgenGrib} or on their 
perturbative~\cite{JAP0405,Claude:FP} and OPE non-perturbative~\cite{oursOPE} 
descriptions. An unambiguous consensus from LQCD, after all this work, pointed that 
$F^2(q^2) G(q^2) \to 0$ when $q \to 0$ (see, for instance, \cite{IlgenGrib,Boucaud:2005ce}) 
and, consequently, that the solution I is excluded provided that the finite-volume 
artefacts are indeed under control. As a matter of the fact, 
finite-volume lattice simulations all agree on a ghost propagator pretty close to that at 
tree-level ($\alpha_F \simeq 0$) and a gluon propagator not far from being a constant at 
vanishing momentum ($\alpha_G \simeq 1$). 

Very recentely~\cite{Bogolubsky:2007ud,Cucchieri:2007md}, simulations on large 
volumes lattices (with a fair control over the finite-volume lattice artifacts) 
yielded solutions for the ghost dressing function confirming that $\alpha_F$ 
is indeed in the vicinity of $0$. Let us now briefly comment about the ghost propagator 
results from these two papers:

\begin{itemize}
\item The authors of ref.~\cite{Bogolubsky:2007ud} simulated the ghost propagator in 
$56^4, 64^4, 72^4, 80^4$ volumes with an impressive control of the finite-size effects
over a huge momentum range from $q^2 \simeq 0.01$ GeV$^2$ 
to $q^2 \simeq 10$ GeV$^2$. They fit an IR exponent, $\alpha_F=-0.174$, 
that appears to be in the vicinity of zero (but negative) and at least much larger that the 
most frequently advocated value ($\simeq - 0.5$). The fit is however 
delicate because the power behavior is dominant, if ever, only on a very small 
momentum domain~\footnote{The fitted IR exponent is unstable, lying more and more in the 
vicinity of zero as the momentum domain becomes smaller 
(see Fig.~2 of ref.~\cite{Bogolubsky:2007ud}).}. Indeed, it is adviseable to try a fitting 
function inspired from \eq{solghostlog}.
Moreover, the numerical solution 
(type II), obtained in ref.~\cite{Boucaud:2008ji}, after a rescaling because of 
the MOM renormalisation, describes strikingly well the lattice ghost propagator data 
from ref.~\cite{Bogolubsky:2007ud} over a large momentum window, from 0.05 GeV to 3 GeV.

\item The authors of ref.~\cite{Cucchieri:2007md} computed an IR ghost propagator exponent, 
$a_G(=-\alpha_F)$, for several 3-dimensional and 4-dimensional lattice volumes (ranging 
from $140^3$ to $320^3$ and from $48^4$ to $128^4$) and collected the results in their 
table 1. The values of $\alpha_F$ from that table are not only in the vicinity of zero 
(although being negative) but they approach systematically zero when the volume 
increases. They fit the power behaviour on a small domain with two or 
four momentum data.

\end{itemize}

 In ref.~\cite{Boucaud:2008ji}, we showed that the $k^2 \log{(k^2)}$ term given 
by \eq{FIR-lat} describes very well the behaviour of a numerical solution of the GPDSE, 
\eq{SD1}, for $\widetilde{g}^2(\mu=1.5~\rm{GeV})=29$ (such a value corresponds to the best description 
of our ghost propagator lattice data) and with a gluon dressing function taken from a
lattice simulation (see Fig.~1 of ref.~\cite{Boucaud:2008ji}). We showed at the same time  
that including a logarithmic divergence 
changes appreciably neither the deduced ghost propagator nor the conclusions about the infrared solutions.

In this same work we analysed in detail the behaviour of the 
numerical solutions of the GPDSE as functions of $\widetilde{g}^2(\mu=1.5~\rm{GeV})$ and discovered that 
a singular solution, behaving as $1/q$ for small momentum (as the relation $2 \alpha_F+\alpha_G=0$ 
requires), appeared only for the specific value $\widetilde{g}^2(\mu=1.5~\rm{GeV})=33.198....$ . This solution 
belongs evidently to what is referred to above as class I, with $\alpha_F = -1/2$ and does satisfy the 
relation $ 2 \alpha_F + \alpha_G = 0$. Furthermore, the closer  
$\widetilde{g}^2(\mu^2)$ to this critical value, the smaller the region near $q=0$ where \eq{FIR-lat} is valid.


\Section{Discussion and Conclusions}
\alinea Thus the present analytical considerations and the previous numerical study converge towards a consistent description of the set of solutions of the ghost Dyson-Schwinger equations:
\begin{itemize}
\item[$\bullet $] A  class of solutions where the ghost dressing function is finite and non zero at $q^2 = 0$ ({\it i.e.} $\alpha_F = 0$), depending continuously on the coupling constant (or equivalently on $F(0)$). Those solutions do not fulfill the relation   $ 2 \alpha_F + \alpha_G = 0$ but appear, for an appropriate value of the coupling, to be in very good agreement with the lattice results.
\item[$\bullet $] An exceptional solution, obtained for a critical value of the coupling is IR-divergent with $\alpha_F = -1/2$. Contrary to the previous ones it satisfies $ 2 \alpha_F + \alpha_G = 0$ but is in clear disagreement with the lattice data over a large range of momenta.
\end{itemize}

We have demonstrated that the discrepancy between LQCD results 
(implying unambiguously that $\alpha_F \simeq 0$ and $\alpha_G \simeq 1$) 
and the usual DSE solutions ($2 \alpha_F+\alpha_G=0$) can 
be solved if the second type (II) of solutions is considered. The existence of this second class 
besides the usual solution (type I) has been proven after
carefully renormalising the GPDSE and  applying a substraction procedure 
to deal with the remaining (after renormalisation) UV singularity. This new solution  yields a finite ghost dressing function at vanishing momentum while $F^2(q^2) G(q^2)$ goes to a zero  when $q \to 0$ contrary to what occurs with type I.

For this (type II) solution, 
an asymptotic formula of the ghost dressing function is obtained that only 
depends on the IR one for the gluon which is taken as an {\it ansatz} 
in this exercise and on the renormalized coupling.
The numerical analysis of the GPDSE in ref.~\cite{Boucaud:2008ji} proves 
that the type II solution exists for any coupling below a given
critical value and that it verifies the asymptotic formula.

The WSTI involving the 3-gluon and the ghost-gluon vertex is particularly useful 
to gain some knowledge about the ghost dressing function:  by simply assuming 
the regularity of some of the tensorial components of the ghost-gluon vertex, 
one can conclude that the ghost dressing function is finite and non zero at vanishing 
momentun. Then, WSTI with the mentioned regularity assumption will discard 
the solution of type I.

Furthermore, LQCD data point to $2 \alpha_F + \alpha_G \simeq 1$ (certainly larger than 0).  
Would one wish to reconcile these data with type I solution ($2 \alpha_F + \alpha_G = 0$), 
very strong finite-volume artifacts would be needed.
 Such a finite-size effect should strengthen the divergence of a ghost propagator 
behaving at finite-volume like at tree-level and damp to zero the gluon propagator. 
This is very doubtful considering that sizeable discrepancies between lattice and solution I 
appear at momenta of the order of $\simeq 0.3$ GeV. On the contrary, 
very recent LQCD data in large volumes
\cite{Cucchieri:2007md, Bogolubsky:2007ud} show a fair stability as the volume increases and, 
if any, a trend towards solution II ($\alpha_F=0$). This is confirmed by 
the numerical analysis of ref.~\cite{Boucaud:2008ji} which proves that
both type I and II solutions {\it live} at infinite volume for different values
of the coupling constant.

It is worth also pointing that 
some attempts to accomodate lattice data within DS coupled equations~\cite{Aguilar:2004sw} and within 
the Gribov-Zwanziger approach~\cite{Dudal:2007cw} led to solutions for gluon and ghost propagators 
that behave pretty much like our solution II does.

Altogether we strongly believe that the question of the ghost propagator
behaviour at small momentum is essentially solved. The solution type II of GPDSE avoiding 
the previous discrepancies, the three methods (DSE, WSTI and LQCD) strikingly converge 
to the same result: {\bf a finite ghost dressing function at vanishing momentum}. 
The case of the gluon
propagator needs further study.

\section*{Acknowledgements}

We are particularly indebted to A.Y. Lokhov and C. Roiesnel for 
partially inspiring this work at their preliminar stages.
This work was done with the help of the Projet ANR-NT05-3\_43577
which is a non thematic project named QCDNEXT, and that of
FPA2006-13825 which is a project from the spanish Ministry of
Science.

\appendix

\section{The Dyson-Schwinger equation as a Ward-Slavnov-Taylor identity}
\label{App}


\alinea A very general method to derive Ward-Slavnov-Taylor identities consists in taking advantage 
of the transformation properties of 

\beq\label{action}
e^{G(J)} = \int {\cal D} (A) \det {\cal M} \exp\left [ i \int d^4x
\left({\cal L} - \frac 1 {2\alpha} (\partial_\mu A_\mu^a)(\partial_\mu A_\mu^a) + 
J_\mu^a A_a^\mu\right)\right]\label{pathInt}
\eeq 
under gauge transformation (cf. \cite{IZ}).

${\cal M}$ is the Faddeev-Popov operator and the notation   $<$, $>_J$  indicates that the source term $J$ has to be kept, although it will eventually be set to 0 (this is denoted in the following by the supression of the $J$ subscript). Taking the derivative of the gauge transformed of eq.~(\ref{pathInt}) with respect to the gauge parameters leads to the general Slavov-Taylor equation

\beq\label{ST}
\frac 1 \alpha <(\partial_\mu A_\mu^a(x))>_J = <\int d^4y 
J_\mu^c(y) \,D_\mu^{cb}(y) F^{(2)ba}(y,x) >_J.
\eeq

$F^{(2)ba}(y,x)$ is the ghost propagator and its presence here is simply due to its very definition as the inverse of the Faddeev-Popov operator.
If one derives eq.~(\ref{ST}) with respect to $J_\rho^d(z)$  one gets  : 
 \bea\label{dJ1}
 \frac 1 \alpha <(\partial_\mu A_\mu^a(x)) A_\rho^d(z)>_J
 &=& <D_\rho^{db}(z)F^{(2)ba}(z,x)>_J \nonumber \\
 &+&  <\int d^4y  J_\mu^c(y)\,D_\mu^{cb}(y) F^{(2)ba}(y,x) A_\rho^d(z)>_J
\eea

A first consequence of this relation is the triviality of the longitudinal gluon propagator. To see this, it suffices to derive both its sides with respect to $z_\rho$ andto set $J$ to zero.
The result is
  \bea
  \frac 1 \alpha <(\partial_\mu A_\mu^a(x))(\partial_\rho A_\rho^d(z))> 
  &=& <\partial_\rho D_\rho^{db}(z)\,F^{(2)ba}(z,x)> \nonumber \\
  &=& \delta_{ad}\,\delta_4(z-x) 
  \eea
 To derive the second line we have invoked the fact that  $\partial_\rho D_\rho^{db}(z)$, the Faddeev-Popov operator, is the inverse of the ghost propagator $F^{(2)}$.
Thus,  in momentum space, the general form of the gluon  propagator for an arbitrary covariant gauge reads 
 \beq\label{G2}
 G^{(2)ab}_{\mu\nu}{(q)} =\delta^{ab}\left[ G^{(2)}(q^2)
 \left(\delta_{\mu\nu}-\frac{q_\mu q_\nu}{q^2}
 \right) + \alpha \frac{q_\mu q_\nu}{(q^2)^2}\right]
 \eeq

Turning back to eq.~(\ref{dJ1}) and setting $J$ go to zero we obtain

\bea\label{dJ10}
  \frac 1  \alpha <(\partial_\mu A_\mu^a(x)) A_\rho^d(z)>
 = <D_\rho^{db}(z)F^{(2)ba}(z,x)> 
 \eea
which  is nothing else than the  GPDSE.
Actually  its l.h.s. involves only the longitudinal part of the gluon propagator, that we have just seeen to be  trivial :
\bea\label{proj}
 \frac 1 \alpha <(\partial_\mu A_\mu^a(x)) A_\rho^d(z)> =
 \partial_\rho \square^{-1}(x,z)
\eea 
 
 As for the r.h.s it can be rewritten as :
 \bea\label{drho}
<D_\rho^{db}(z)F^{(2)ba}(z,x)> =<\partial_\rho F^{(2)da}(z,x)> + 
i<g f^{deb} A_\rho^{e}(z) F^{(2)ba}(z,x)>
 \eea
 
 The 3-point gluon-ghost Green's function can be expressed in terms of vertex functions and propagators through
 \bea\label{G3}
 &\widetilde G^{(3)fgh}_\rho(p,q,r) \equiv -i\int d^4xd^4td^4 z e^{ipx}e^{irz}e^{iqt} 
 <A_\rho^{f}(t) F^{(2)gh}(z,x)> \\
 &= g \frac {F(p^2)}{p^2} \frac{F(r^2)}{r^2}  
 \left[ \frac{G(q^2)}{q^2}
 \left(\delta_{\rho\nu}-\frac{q_\rho q_\nu}{q^2}
 \right) + \alpha \frac{q_\rho q_\nu}{(q^2)^2}\right]f^{fgh}
 \widetilde\Gamma_\nu(p,r;q)(2\pi)^4\delta_4(p+q+r) \nonumber
 \eea
 
 We Fourier transform \eq{dJ10}
, use  equations~(\ref{proj}), (\ref{drho}) and
  (\eq{G3})
and obtain
 
 \bea\label{dJ10f}
 \frac{k_\rho}{k^2} &=&  \frac{k_\rho}{k^2} F(k^2) - g 
 f^{deb}f^{eba}\int \frac{d^4q}{(2\pi)^4}
 \frac{F(k^2)}{k^2} \frac{F((k+q)^2)}{(k+q)^2} \nonumber\\
& &  \left[ \frac{G(q^2)}{q^2}
 \left(\delta_{\rho\nu}-\frac{q_\rho q_\nu}{q^2}
 \right) +  \alpha \frac{q_\rho q_\nu}{(q^2)^2}\right]
 \widetilde\Gamma_{\nu}(k,-k-q;q)
 \eea

 The usual form is recovered by multiplying with  $k_\rho$
and dividing by $F(k^2)$, which leads to  
  \bea\label{GPDSE}
 F^{-1}(k^2) &=& 1 - g f^{deb}f^{eba}\int \frac{d^4q}{(2\pi)^4}
 \frac{F((k+q)^2)}{(k+q)^2}\nonumber\\
 &\,&\left[ \frac{G(q^2)}{q^2}\left(k_\nu-\frac{(qk) q_\nu}{q^2}\right) +  \alpha \frac{(qk) q_\nu}{(q^2)^2}\right]
 \widetilde\Gamma_{\nu}(k,-k-q;q)
 \eea

 This is a  general result, valid in any covariant gauge. Of course  the $\alpha$ depending (longitudinal) term disappear  in Landau gauge.
 
 $\widetilde\Gamma_{\nu}(k,-k-q;q)$ is related to the quantity previously introduced in section~\ref{SlavTayl} through the relation
 
$$ \widetilde\Gamma_{\nu}(k,-k-q;q)=-i g k_\mu\widetilde\Gamma_{\mu\nu}(k,-k-q;q)$$
 
 and is usually decomposed into
$\widetilde\Gamma_{\nu}(k,-k-q;q)= g\left[ k_\nu H_1(k,q)+   q_\nu H_2(k,q)\right]  $.

Inserting this in eq.(\ref{GPDSE}) and restricting to the Landau gauge case gives 
\bea\label{GPDSEL}
 F^{-1}(k^2) = 1 +g^2 N_c \int \frac{d^4q}{(2\pi)^4}
 \frac{F((k+q)^2)}{(k+q)^2}\left[ \frac{G(q^2)}{q^2}\left(\frac{(qk)^2}{q^2} -k^2\right) \right]H_1(k,q)
 \eea




\vspace{2cm}
\addcontentsline{toc}{section}{References}
\begin{thebibliography}{99}

\bibitem{Alkofer:2000wg}
  R.~Alkofer and L.~von Smekal,
  Phys.\ Rept.\  {\bf 353} (2001) 281
  [arXiv:hep-ph/0007355].


\bibitem{IlgenGrib}
 A.~Sternbeck, E.-M.~Ilgenfritz, M.~Müller-Preussker and A.~Schiller,
Nucl.\ Phys.\ Proc.\ Suppl.\  {\bf 140} (2005) 653;
  AIP Conference Proceedings {\bf 756} (2005) 284,
  [arXiv:hep-lat/0412011].

\bibitem{Boucaud:2005ce}
  P.~Boucaud {\it et al.},
 [arXiv:hep-ph/0507104 ].

\bibitem{Alkofer:2008jy}
R.~Alkofer, M.~Q.~Huber and K.~Schwenzer,
arXiv:0801.2762 [hep-th].


\bibitem{Finite:2006}
  Ph.~Boucaud {\it et al.},
  JHEP {\bf 0606} (2006) 001
  [arXiv:hep-ph/0604056].

\bibitem{Boucaud:2007va}
  Ph.~Boucaud {\it et al.},
  Eur.\ Phys.\ J.\  A {\bf 31} (2007) 750
  [arXiv:hep-ph/0701114].


\bibitem{Boucaud:2008ji}
  Ph.~Boucaud, J.~P.~Leroy, A.~L.~Yaouanc, J.~Micheli, O.~Pene and J.~Rodriguez--Quintero,
 arXiv:0801.2721 [hep-ph].

\bibitem{ours-ST}
  Ph.~Boucaud {\it et al.},
  JHEP {\bf 0703} (2007) 076
  [arXiv:hep-ph/0702092].

\bibitem{LvS}
C.~Lerche and L.~von Smekal  Phys.Rev.D65:125006,2002.
[arXiv:hep-ph/0202194]



\bibitem{Adelaide}
 F. ~D.~ R.~ Bonnet, P. ~O. ~Bowman, D.  ~B. ~Leinweber, A. ~G. ~Williams, J. ~M. ~Zanotti,
Phys.\ Rev.\ D {\bf64} (2001) 034501
[arXiv: hep-lat/0101013]
  
\bibitem{Bloch:2003yu}
  J.~C.~R.~Bloch,
  Few Body Syst.\  {\bf 33} (2003) 111
  [arXiv:hep-ph/0303125].
 
 \bibitem{Taylor} J. ~C. ~Taylor,
 Nuclear Physics B
 Volume 33, Issue 2 , 1 November 1971, Pages 436-444   
\\
  A.~A.~Slavnov,
  Theor.\ Math.\ Phys.\  {\bf 10} (1972) 99
  [Teor.\ Mat.\ Fiz.\  {\bf 10} (1972) 153].

\bibitem{Ball:1980ax}
James~S. Ball and Ting-Wai Chiu.
\newblock {\em Phys. Rev.}, D22:2550, 1980.
\newblock ERRATUM ibid 23(1981),3805.

\bibitem{Chetyrkin:2000dq}
  K.~G.~Chetyrkin and A.~Retey,
 [arXiv:hep-ph/0007088].

\bibitem{Bogolubsky:2007ud}
 I.~L.~Bogolubsky, E.~M.~Ilgenfritz, M.~Muller-Preussker and A.~Sternbeck,
 arXiv:0710.1968 [hep-lat].


\bibitem{Cucchieri:2007md}
 A.~Cucchieri and T.~Mendes,  
  arXiv:0710.0412 [hep-lat].

%
\bibitem{ZWA94}
    D.~Zwanziger, Nucl.~Phys.~B412 (1994) 657.
%
\bibitem{SUM96}
    H.~Suman, K.~Schilling, Phys.~Lett.~B373 (1996) 314, 
    arXiv:hep-lat/95120003.
%
\bibitem{JAP0405}
    S.~Furui, H. Nakajima, Phys.~Rev.~D69 (2004) 074505, 
    [arXiv:hep-lat/0305010];
    S.~Furui, H. Nakajima, Phys.~Rev.~D70 (2004) 094504, 
    [arXiv:hep-lat/0403021].
%
\bibitem{Claude:FP}
 Ph.~Boucaud,  J.P.~Leroy,  A.~Le~Yaouanc,  A.Y.~Lokhov ,
J. Micheli,  O. P\`ene,  J.~Rodr\'iguez-Quintero  and 
C.~Roiesnel,
 [arXiv:hep-lat/0506031]

%
\bibitem{oursOPE}
  Ph.~Boucaud {\it et al.},
  JHEP {\bf 0601} (2006) 037
  [arXiv:hep-lat/0507005].



\bibitem{Davydychev:1996pb}
A.~I.~Davydychev, P.~Osland and O.~V.~Tarasov,
Phys.\ Rev.\ D {\bf 54}, 4087 (1996)
[Erratum-ibid.\ D {\bf 59}, 109901 (1999)]
[arXiv:hep-ph/9605348].


  

\bibitem{Aguilar:2004sw}
  A.~C.~Aguilar and A.~A.~Natale,
  JHEP {\bf 0408} (2004) 057;
  A.~C.~Aguilar and J.~Papavassiliou,
  JHEP {\bf 0612} (2006) 012;
  Eur.\ Phys.\ J.\  A {\bf 31} (2007) 742.


\bibitem{Dudal:2007cw}
  D.~Dudal, S.~P.~Sorella, N.~Vandersickel and H.~Verschelde,
  arXiv:0711.4496 [hep-th].

\bibitem{IZ}
  C. Itzykson and J.-B. Zuber, Quantum Field Theory, McGraw-Hill ed. (1980) pp 594 sqq.
  arXiv:0711.4496 [hep-th].


\end{thebibliography}
\end{document}